\documentclass[sigconf, noacm, pbalance]{acmart}

\AtBeginDocument{%
  }
\usepackage{amsmath,amsfonts}
\usepackage[ruled,vlined,linesnumbered]{algorithm2e}
\usepackage{graphicx}
\usepackage{textcomp}
\usepackage{listings}
\usepackage{xcolor}
\usepackage[english]{babel}
\usepackage{color,colortbl}
\usepackage{xspace}
\usepackage{multirow}
\usepackage{booktabs}
\usepackage{url}
\usepackage{xurl}
\usepackage{tipa}
\usepackage[T1]{fontenc}
\usepackage[utf8]{inputenc}
\usepackage{threeparttable}
\usepackage{placeins}
\usepackage{pifont}
\usepackage{tabularx}
\usepackage[printonlyused]{acronym}
\usepackage[most]{tcolorbox}
\usepackage{multicol}
\usepackage{wrapfig}
\usepackage{adjustbox}

\usepackage{hyperref}
\hypersetup{
	pdftitle={},
	pdfauthor={},
	pdfsubject={},
	pdfkeywords={},
	bookmarksnumbered=true,
	bookmarksopen=false,
	colorlinks=true,
	pdfstartview=FitB,
	pdfpagemode=UseOutlines
}

\usepackage{tikz}
\usetikzlibrary{shapes,arrows,positioning,fit}
\usetikzlibrary{backgrounds}
\usetikzlibrary{arrows.meta, positioning, calc}

\definecolor{mygreen}{rgb}{0,0.6,0}
\definecolor{mygray}{rgb}{0.5,0.5,0.5}
\definecolor{mymauve}{rgb}{0.58,0,0.82}
\definecolor{pastelcyan}{rgb}{0.741, 0.867, 0.894}
\definecolor{skyice}{rgb}{0.62, 0.776, 0.953} 
\definecolor{warmlinen}{rgb}{0.949, 0.937, 0.906}
\definecolor{powderblue}{RGB}{152,193,217}
\definecolor{sageleaf}{RGB}{233, 237, 201}
\definecolor{blushcloud}{RGB}{241,231,231}
\definecolor{gainsboro}{RGB}{220,220,220}

\definecolor{mygold}{RGB}{175, 149, 0}   
\definecolor{mysilver}{RGB}{180, 180, 180} 
\definecolor{mybronze}{RGB}{173, 138, 86} 

\usepackage[font=footnotesize,skip=0pt,belowskip=0pt,aboveskip=0pt]{caption}
\usepackage{caption}
\usepackage{subcaption}

\newcommand{\squishlist}{
	\begin{list}{$\bullet$} {
			\setlength{\itemsep}{0pt}
			\setlength{\parsep}{0pt}
			\setlength{\topsep}{0pt}
			\setlength{\partopsep}{0pt}
			\setlength{\leftmargin}{1.0em}
			\setlength{\labelwidth}{1em}
			\setlength{\labelsep}{0.5em}
		}
		}
		\newcommand{\squishend}{
	\end{list}
}

\newcommand{\cut}[1] {}

\cut{\zhiyun{a few things we should do before submission:
		\begin{itemize}
			\item check captilization consistency, ``figure -> Figure'', caption of a figure/table should have the first letter captilized only
			\item center the caption of all figures/tables.
			\item remove the ``period'' mark in captions.
			\item spell check / grammar check (put all sentences in grammarly)
			\item for all the functions referened in textit\{\}, we should add the brackets. For example, ``textit\{fun\} -> textit\{fun()\}''
			\item global replace of ``\ textit'' with ''\ texttt''. Also, there is no need to add the quotation marks around \texttt.
			\item check the text that references the figure and the figure itself are always co-located on the same page (or one page before/after).
		\end{itemize}
	}}

\setcopyright{none}
\renewcommand\footnotetextcopyrightpermission[1]{ 
  \footnotetext{
  This work was performed when the first author was 
  interned at Meta.
}}

\copyrightyear{2018}
\acmYear{2018}
\acmDOI{XXXXXXX.XXXXXXX}
\acmConference[Arxiv Preprint]{}{Dec 2025}{Earth}

\usepackage{hyperref}

\newcommand{\work}{\mbox{\textsc{TritonForge}}\xspace}

\author[Haonan Li et al.]{Haonan Li$^{1}$, Keyu Man$^{2}$, Partha Kanuparthy$^{2}$, Hanning Chen$^{3}$, Wei Sun$^{2}$, Sreen Tallam$^{2}$, Chenguang Zhu$^{2}$, Kevin Zhu$^{2}$, Zhiyun Qian$^{1}$}

\email{{hli333, zhiyunq}@ucr.edu, {kman, pka, wesun, sreen, cgzhu, kevzh}@meta.com,
hanningc@uci.edu}

\affiliation{%
  $^1$\textit{University of California, Riverside}
  $^2$\textit{Meta}
  $^3$\textit{University of California, Irvine} \\
  \country{}
}

\begin{document}

\title{\work: Profiling-Guided Framework for Automated Triton Kernel Optimization}



\begin{abstract}
    High-performance GPU kernel optimization remains a critical yet
    labor-intensive task in modern machine learning workloads. Although Triton,
    a domain-specific language for GPU programming, enables developers to write
    efficient kernels with concise code, achieving expert-level performance
    still requires deep understanding of GPU architectures and low-level
    performance trade-offs. We present \work, a profiling-guided framework for
    automated Triton kernel optimization. \work integrates kernel analysis,
    runtime profiling, and iterative code transformation to streamline the
    optimization process. By incorporating feedback from profiling
    results, the system identifies performance bottlenecks, proposes targeted
    code modifications, and evaluates their impact automatically. Across diverse
    kernel types, \work achieves up to 5x performance
    improvement over baseline implementations and on average 1.76x of the cases
    are successful, providing a foundation for future research in automated GPU
    performance optimization.
\end{abstract}



\settopmatter{printfolios=true}

\maketitle

\section{Introduction}

High-performance GPU computing has become the cornerstone of modern machine
learning and scientific computing \cite{NVBPG2025}. As deep learning models grow
increasingly complex and datasets expand to unprecedented scales
\cite{Kaplan2020,Chinchilla2022}, the demand for efficient GPU kernels has never
been greater. Triton, adopted as the default backend compiler for PyTorch 2.0
\cite{PyTorch2_2024}, has emerged as a powerful domain-specific language (DSL)
for writing high-performance kernels that can rival hand-optimized CUDA code
\cite{TritonMAPL2019,OpenAI_TritonBlog}. Its block-based programming model
allows developers to express complex tensor operations without explicitly
managing thread synchronization or shared memory allocation.


However, despite its syntactic simplicity, achieving peak performance in Triton
remains a non-trivial challenge that demands deep hardware awareness. While
Triton abstracts away the thread-level details of CUDA, it shifts the
optimization burden to tile-level data flow management
\cite{FlashAttention2022}. Developers must manually tune critical
meta-parameters, such as block sizes (e.g., \texttt{BLOCK\_M},
\texttt{BLOCK\_N}), the number of pipeline stages (\texttt{num\_stages}) and
warp configurations (\texttt{num\_warps}) to match the target GPU's L2 cache
size and register file limits \cite{TritonTutorials}. A suboptimal
configuration, such as a tile size that causes cache thrashing or memory bank
conflicts, can degrade performance by an order of magnitude even if the code is
functionally correct. Furthermore, the rapid evolution of GPU architectures
implies that optimization strategies must be continuously adapted, making manual
tuning an ongoing and labor-intensive burden \cite{AutoTriton2025,
AnsorOSDI2020}.

Recent advances in Large Language Models (LLMs) have demonstrated remarkable
capabilities in code generation, optimization, and code understanding
\cite{KernelBench2024,CUDA_LLM_2025}. These models can understand complex
programming patterns, suggest optimizations, and even generate high-performance
code from high-level specifications. This presents an
opportunity to democratize GPU kernel optimization by leveraging LLMs to
automate and accelerate the optimization process.


Nevertheless, the optimization of Triton kernels presents several critical challenges that
limit the widespread adoption of high-performance GPU computing
\cite{TritonMAPL2019,TritonPerf2023}. First, the complexity of underlying GPU
architectures requires developers to understand memory hierarchies, parallel
execution models, and performance-critical scheduling decisions
\cite{TritonLangGuide,tritonAutotune}. This expertise barrier prevents many
developers from achieving optimal performance, resulting in kernels that fail to
fully utilize modern GPUs \cite{TritonPerf2023}.

Second, the manual optimization process is inherently iterative and
time-consuming \cite{TritonLangGuide,NsightProfiling2024}. Developers must
repeatedly profile, analyze, and tune kernels to identify performance
bottlenecks, requiring extensive experimentation with tiling, parallelism
parameters, and memory layouts \cite{tritonConfigAPI,OpenAI_TritonBlog}. Such
trial-and-error search not only delays development cycles but may miss optimal
implementations due to the large search space \cite{AnsorOSDI2020}.

Third, the rapid evolution of GPU architectures creates a moving target for
optimization. Techniques and parameter choices that
perform well on one architecture (e.g., Ampere) may underperform on newer ones
(e.g., Hopper), requiring continuous retuning to achieve peak performance
\cite{TritonPerf2023}. This hinders kernel portability and
long-term maintainability.

Finally, the lack of automated, end-to-end tooling for Triton performance
optimization means that optimization knowledge remains siloed among expert
practitioners \cite{tritonAutotune,OpenAI_TritonBlog}. Even though Triton
simplifies GPU programming, developers must still manually discover performant
configurations \cite{tritonConfigAPI}. An ideal solution would provide
intelligent automation that captures expert knowledge and adapts to
hardware-specific performance patterns \cite{KernelBench2024,AnsorOSDI2020}.



This work introduces \work, a practical framework for LLM-assisted Triton kernel
optimization that integrates performance profiling feedback directly into the
code generation process. 
Our contributions are as follows:

\squishlist
\item \textbf{LLM-Guided Kernel Generation and Optimization.}
We propose a unified workflow where Large Language Models generate Triton kernel
variants and iteratively refine them using structured performance feedback from
Nsight Compute. By closing the loop between profiling and code generation,
\work enables optimization without exhaustive manual tuning.

\item \textbf{Profiling-Aware Code Improvements.}
Our method translates low-level profiling metrics (e.g., memory throughput,
warp occupancy, instruction stalls) into actionable code transformations.
This allows the LLM to reason over hardware-specific performance constraints
while preserving semantic correctness.

\item \textbf{Automated Performance Evaluation Pipeline.}
We build a fully automated benchmarking and analysis system that executes
kernels, extracts bottleneck indicators, and provides quantitative feedback to
guide optimization iterations—reducing reliance on manual profiling expertise.

\item \textbf{Empirical Demonstration of Effectiveness.}
We evaluate \work on representative Triton kernels, demonstrating performance
improvements of up to 5x over baseline implementations and on average 1.76x of the cases are successful.
 These results show that integrating
profiling signals into the code generation process accelerates convergence
toward high-performance implementations.

\squishend

\section{Background}

\begin{figure*}[h!]
    \centering
    \includegraphics[width=0.97\textwidth]{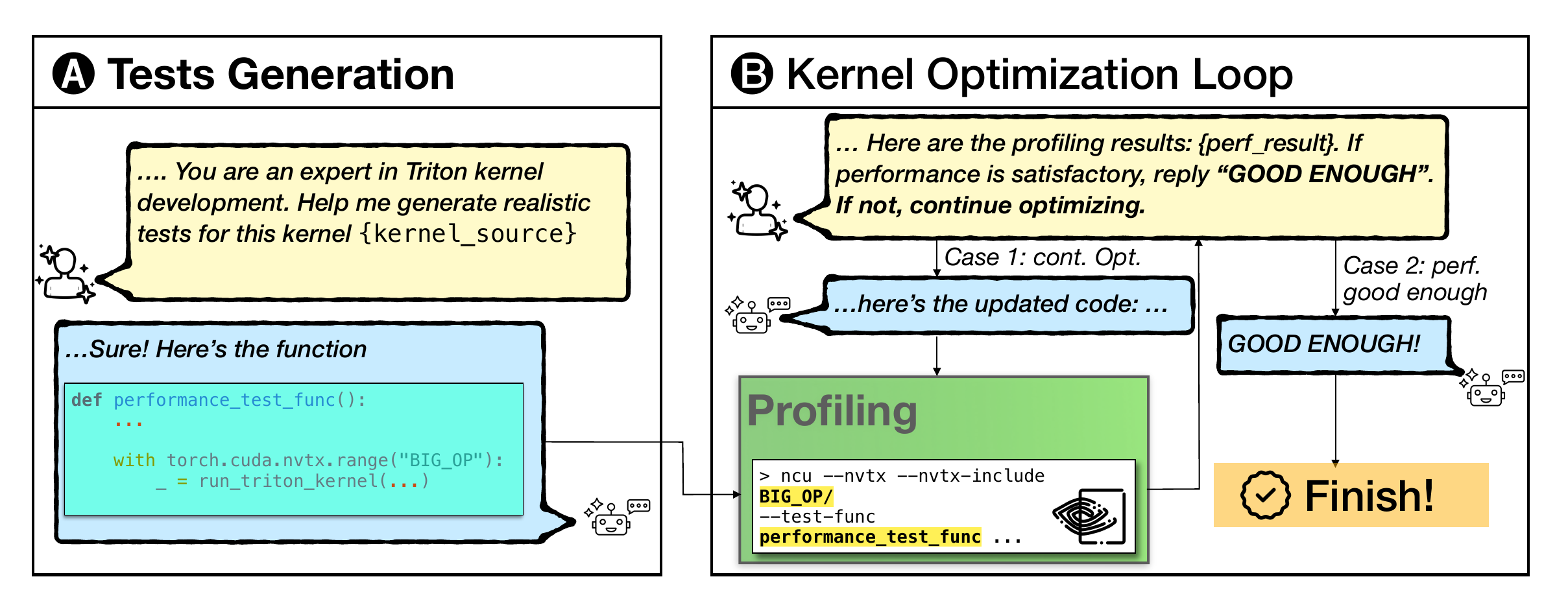}
    \caption{Overview of \work.
    The prompts and responses shown in the figure
     are simplified for clarity.
    }
    \label{fig:methodology}
\end{figure*}

This section presents the background knowledge of our work: Triton and GPU
kernel optimization, and LLM-based code generation.

\vspace{0.5em}
\noindent
\textbf{Triton Framework and Design Philosophy.} 
Triton~\cite{OpenAI_TritonBlog} is an open-source GPU programming framework introduced by
OpenAI, designed to simplify high-performance kernel development through a
Python-embedded domain-specific language. By abstracting many low-level CUDA
details while still exposing explicit control over memory hierarchy and parallel
execution, Triton enables developers to write succinct kernels that can reach
performance comparable to expert-written CUDA implementations. Its tile-based
programming model, programmatic memory movement, and auto-generated
vectorization make it particularly suitable for machine learning workloads.

In recent years, Triton has seen rapid adoption in both academia and industry,
especially following its integration into PyTorch 2.0 as the default backend of
TorchInductor\cite{torchinductor2022,PyTorch2_2024}. Many modern LLM training and
inference systems rely on Triton for implementing fused operators,
memory-efficient attention kernels, softmax variants, and other
performance-critical GPU primitives \cite{xformers, dao2024flashattention2}. This highlights
 Triton as a key enabling technology for
efficient large-scale training, demonstrating significant speedups over
traditional CUDA baselines.

The growing prevalence of Triton in practical deployments—ranging from operator
fusion in production inference servers to custom kernels for state-of-the-art
LLM architectures—underscores the importance of rigorously understanding its
programming abstractions and optimization behaviors. As Triton becomes a central
component of the modern deep learning software stack, systematic analysis of
kernel performance, memory access patterns, and compiler optimization strategies
provides valuable insight for both practitioners and researchers.

\vspace{0.5em}
\noindent
\textbf{Current Optimization Approaches.} Traditional Triton optimization relies
heavily on manual tuning of key parameters such as block sizes, memory access
patterns, and loop unrolling strategies.  The introduction of Triton
2.0~\cite{nvidia_blackwell_2025} has brought significant improvements in compilation
efficiency and optimization capabilities, while recent work on automated Triton
kernel generation~\cite{liger_kernel_2025} has shown promise in reducing manual
optimization effort. However, these approaches often require domain expertise
and extensive experimentation.


\vspace{0.5em}
\noindent
\textbf{GPU Kernel Optimization.} 
GPU kernel optimization has traditionally relied on auto-tuning and heuristic
search techniques to explore kernel parameters and memory layouts efficiently
\cite{bergstra2022benchmarking,koukos2018kernel}. Compiler-driven and
rule-based approaches further improve performance through static analysis and
structured code transformations \cite{lou2024automatic}. More recently,
LLM-assisted methods have shown strong potential for GPU kernel generation and
performance optimization, particularly in PyTorch and CUDA environments
\cite{KernelBench2024,chen2025cuda}. However, these efforts
overlook Triton, where optimization expertise remains a barrier. \work
addresses this gap by integrating profiling-aware feedback into the
Triton kernel optimization process.

\vspace{0.5em}
\noindent
\textbf{LLM for Code Generation.}
Recent research in large language models (LLMs) designed for code generation has
made significant progress \cite{Huynh2025LLMCodeSurvey}. For instance, OpenAI's
Codex agent \cite{OpenAI_Codex_2025} exemplifies the new wave of AI-coding
systems that integrate natural language specification with execution, tool use,
and multi-step reasoning. Anthropic's Claude Sonnet 4.5
\cite{Anthropic_Claude_Sonnet4.5_2025} attains state-of-the-art performance on
software engineering and agentic coding tasks. However, neither of these models
have been developed specifically for the programming model of Triton kernels,
with their unique syntax and semantics. In this work, we propose to integrate
profiling feedback from Nsight Compute into the LLM generation loop to steer the
model toward producing optimized Triton code.

\vspace{0.5em}
\noindent
\textbf{LLM for Performance Optimization.} Recent advancements have expanded the
role of LLMs from functional code generation to
automated performance optimization. For instance, recent
work~\cite{shypula2024learning} introduced the PIE benchmark to train models
specifically for predicting performance-improving edits in C++ programs. To
explore the optimization space more effectively, the search-based framework
SBLLM~\cite{wu2025search} combines LLMs with evolutionary strategies,
outperforming direct prompting methods. Scaling this approach,
PEACE~\cite{ren2025peace} targets project-level efficiency via a hybrid code
editing strategy, rather than focusing solely on isolated functions. Unlike
these general-purpose approaches targeting CPU-bound C++ or Python code,
\textsc{TritonForge} specifically addresses GPU kernel optimization by
leveraging the profiling feedback to guide the LLM.

\section{Methodology}

\subsection{Overview}

Our methodology centers around an LLM agent that integrates with NVIDIA Nsight
profiling tools to provide intelligent, data-driven optimization suggestions for
Triton kernels. The framework operates through a multi-stage pipeline that
combines test generation, performance profiling, and optimization loop to
deliver context-aware improvements.

\vspace{0.5em}
\noindent
\textbf{Core Architecture.} 
As shown in Figure \ref{fig:methodology}, \work consists of two main stages:
(A) Test Generator: An LLM agent that understands Triton code semantics and generates performance
tests and (B) Kernel Optimizer: An LLM agent that reads the performance metrics from NVIDIA Nsight Compute
 and generates optimized Triton code.
Additionally, 
\work also contains another agent called Fault-Aware Remediation Agent,
which is responsible for reviewing the compilation and runtime errors of the generated code and 
trying to fix them if there are any.

\vspace{0.5em}
\noindent
\textbf{Workflow Overview.} 
The workflow of \work is as follows:
(1) Test Generator (A) understands the input Triton code and generates performance
tests.
(2) Profiling module collects performance metrics from the generated performance
tests and identifies bottlenecks.
(3) Kernel Optimizer (B) reads the performance metrics from the profiling module
and generates optimized Triton code.
The (2) and (3) are repeated iteratively until the optimized Triton code is satisfactory.

\subsection{Dataset}

\textbf{TritonBench.} We leverage the TritonBench
dataset~\cite{li-etal-2025-tritonbench}, the first comprehensive benchmark for
Triton operator generation. TritonBench features 184 real-world operators
collected from GitHub repositories (>100 stars), providing a diverse
collection of Triton kernels spanning various computational patterns and
optimization levels. The dataset includes kernels for matrix operations,
convolution operations, attention mechanisms, and custom computational kernels,
each with corresponding unit tests.

\vspace{0.5em}
\noindent
\textbf{Augmented Dataset for Optimization.}
To enable effective LLM-based optimization, we further enrich each kernel in
TritonBench with additional contextual and performance data. Specifically, for
each operator, we generate (1) \emph{performance test cases},
(2) \emph{GPU hardware specifications}, and (3) \emph{profiling traces}.

First, we leverage the existing correctness tests and input-output pairs from
the TritonBench dataset to verify functional equivalence across kernel variants,
while generating our own performance tests to measure execution time under
controlled input sizes. Second, we record
detailed hardware configurations, including GPU model, SM count, clock
frequency, and memory hierarchy, which provide the contextual constraints for
performance prediction and optimization. Third, we collect low-level runtime
metrics using NVIDIA Nsight Compute, capturing information such as memory
bandwidth utilization, instruction throughput, and kernel occupancy.

Together, these additions transform TritonBench into a richer dataset that
captures not only the source code of GPU kernels but also their behavioral and
hardware-dependent characteristics. This augmented dataset enables our model to
reason jointly over code semantics, runtime behavior, and hardware context when
generating optimized Triton implementations.


\subsection{Triton Kernel Optimization Pipeline}

\SetKwInput{KwIn}{Input}
\SetKwInput{KwOut}{Output}
\SetKwFunction{Proposal}{Proposal}
\SetKwFunction{BuildRun}{BuildRun}
\SetKwFunction{Remediation}{Remediation}
\SetKwFunction{Profile}{Profile}
\SetKwFunction{Arbiter}{Arbiter}
\SetKwFunction{RefineHint}{RefineHint}
\SetKwRepeat{Do}{repeat}{until} 

\begin{algorithm}[t]
\DontPrintSemicolon
\caption{\textsc{Profiling-Guided Triton Kernel Optimization}}
\label{alg:triton-llm-loop}

\KwIn{\parbox{\linewidth}{%
\begin{tabular}{@{}ll@{}}
$H$: & hardware profile \\
$K$: & initial Triton kernel \\
$R$: & baseline profiler report \\
\end{tabular}}}
\KwOut{$K^\star$: optimized kernel;\ $R^\star$: profiler report}

$K^\star \gets K$;\ $R^\star \gets R$;\ $dec \gets continue $\;

\Do{dec = finish}{
  $K' \gets$ \Proposal{H, K, R}\;
  $res$ $\gets$ \BuildRun{$K'$}\;
  \While{res = fail}{
    $K' \gets$ \Remediation{$K'$, logs}\;
    $res$ $\gets$ \BuildRun{$K'$}\;
  }
    \Else{
    $R' \gets$ \Profile{$K'$}\;
    $dec \gets$ \Arbiter{$R'$, R$^\star$}\;
    \If{dec = accept}{$K^\star \gets K'$;\ $R^\star \gets R'$}
    $K \gets$ \RefineHint{$K$, $R$, $R'$}\;
    $R \gets R'$\;
  }
}
\Return{$K^\star, R^\star$}\;
\end{algorithm}

\textbf{Model Selection.} We employ Google's Gemini-2.5-Pro for our
optimization framework. These models provide the necessary capabilities for
understanding complex code patterns, reasoning about optimization strategies,
and generating high-quality code.




\vspace{0.5em}
\noindent
\textbf{Test Generation.} As shown in Figure \ref{fig:methodology}, 
(A) Test Generator is used to generate performance tests for the input Triton kernel.
We leverage the existing tests and ask LLM to generate more comprehensive tests 
under specific nvtx range label for profiling (i.e., \texttt{BIG\_OP})

\vspace{0.5em}
\noindent
\textbf{Kernel Optimization Loop.} The detailed optimization loop 
is shown in Algorithm \ref{alg:triton-llm-loop}. (Part (B) in Figure \ref{fig:methodology}).
Specifically, it consists of the following steps:

\squishlist
\item{\textbf{Initialization-and-Proposal.}}
    \emph{Inputs:} device capabilities $C$, kernel source $K$, performance
    report $R$.  
    \emph{Output:} proposal kernel $K'$.  
    \emph{Role:} Generates candidate code refinements conditioned on hardware
    and existing bottlenecks (e.g., tiling/block shapes, \texttt{num\_warps},
    vectorization width, memory layout, prefetching, loop fusion/splitting)
    while preserving correctness constraints.
    
    \item{\textbf{Fault-Aware Remediation.}}
    \emph{Inputs:} kernel $K'$, diagnostic logs.  
    \emph{Output:} repaired kernel $\widetilde{K}$.  
    \emph{Role:} Automatically fixes compilation and runtime failures using
    targeted LLM edits informed by compiler/runtime logs.
    
    \item{\textbf{Performance-Arbiter.}}
    \emph{Inputs:} profiler report $R'$, best-so-far $R^\star$, thresholds
    $\Theta$.  
    \emph{Output:} decision $\in
    \{\texttt{accept},\texttt{continue},\texttt{finish}\}$.  
    \emph{Role:} Multi-criteria decision based on throughput/latency speedup,
    robustness checks, resource limits, and diminishing return criteria.
    
    \item{\textbf{Targeted-Refinement-Hint.}}
    \emph{Inputs:} previous report $R$, new report $R'$.  
    \emph{Output:} refined kernel $K$.  
    \emph{Role:} Converts profiler deltas into actionable feedback (e.g., memory
    stalls $\Rightarrow$ improve tiling or staging, low occupancy $\Rightarrow$
    adjust register pressure).

\squishend
\vspace{0.5em} 
\noindent
It is important to note that the procedure forms a potentially
unbounded loop, as the remediation stage may not always succeed in
resolving compilation or runtime errors. To ensure termination and avoid
infinite execution, we impose a strict upper limit on the number of optimization
iterations (set to 8 in our experiments).

\subsection{Implementation Details}

\textbf{Codebase.} 
The \work codebase comprises approximately 3,000 lines of Python code responsible
for implementing the LLM agent, orchestrating kernel generation, and managing
automated profiling. In addition, the repository contains around 300 lines of
prompt templates and instructions used by the LLM agent throughout the
optimization process.

\vspace{0.5em}
\noindent
\textbf{Execution Time.} 
The overall runtime per kernel depends on the kernel's complexity and size. 
The profiling time is typically within minutes for multiple warmups and iterations.
The
primary performance bottleneck is the LLM inference time, which typically takes
several minutes per optimization query. To improve throughput, LLM requests can
be parallelized using multiple API calls. In our experiments, the end-to-end
optimization process for a single kernel generally completes in approximately 20
minutes.

\vspace{0.5em}
\noindent
\textbf{Profiling.}
We assess performance using both runtime latency and hardware-level profiling.
The kernel execution time is measured using CUDA events. We run three warm-up
iterations to eliminate JIT and frequency ramp-up effects, followed by five
timed iterations. Each iteration uses \texttt{torch.cuda.Event} and an explicit
synchronization to ensure accurate kernel-only timing.

For low-level profiling, we use NVIDIA Nsight Compute 2025.2.1.0 to  collect
hardware-level metrics such as occupancy and memory efficiency. NVTX markers
isolate the Triton kernel region of interest. Profiling is invoked with:
\begin{lstlisting}[language=bash]
ncu -f --nvtx --nvtx-include <range_label> ...
\end{lstlisting}
This procedure enables reproducible timing and detailed insight into GPU
execution characteristics, supporting fair comparison across optimization
variants.

We demonstrate the entire profiling process with a real example 
later in \S\ref{sec:case_study_3}.

\section{Experiments}

This section presents a comprehensive evaluation of \work.
We conduct a series of experiments to evaluate the effectiveness and behavioral characteristics of large language models (LLMs) in optimizing Triton kernels. 
Our study aims to answer the following research questions:

\squishlist
    \item \textbf{RQ1:} \textit{Can LLMs generate optimized kernels that outperform the original implementations in execution speed?}  
    We examine the overall success rate, the magnitude of speedups, and the distribution of improvements across kernel complexity levels.

    \item \textbf{RQ2:} \textit{How do iterative optimization and the number of refinement rounds affect performance outcomes?}  
    We analyze convergence trends and diminishing returns over LLM interaction cycles.

    \item \textbf{RQ3:} \textit{What behavioral patterns emerge from the LLM during kernel optimization?}  
    We investigate changes in code length and verbosity as indirect signals of the model’s reasoning and optimization strategies.

    \item \textbf{RQ4:} \textit{Which components of \work most contribute to performance gains?}  
    We perform an ablation study to isolate the effects of individual modules such as profiling-guided feedback and prompt structure.
\squishend


\begin{figure*}[t!]
    \centering
    \begin{subfigure}[t]{0.495\textwidth}
      \includegraphics[width=\linewidth]{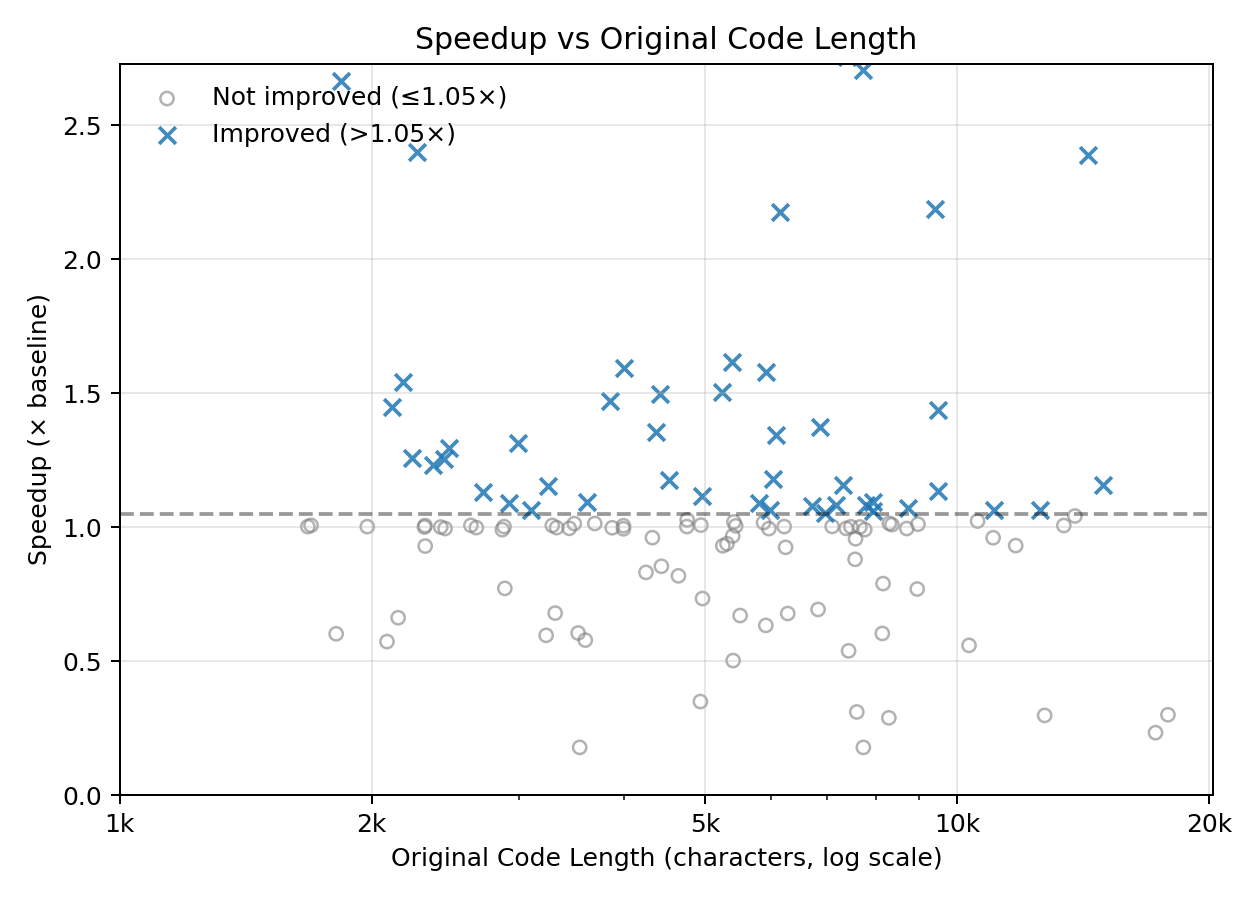}
      \caption{Speedup vs code length.}
    \end{subfigure}
    \hfill
    \begin{subfigure}[t]{0.495\textwidth}
      \includegraphics[width=\linewidth]{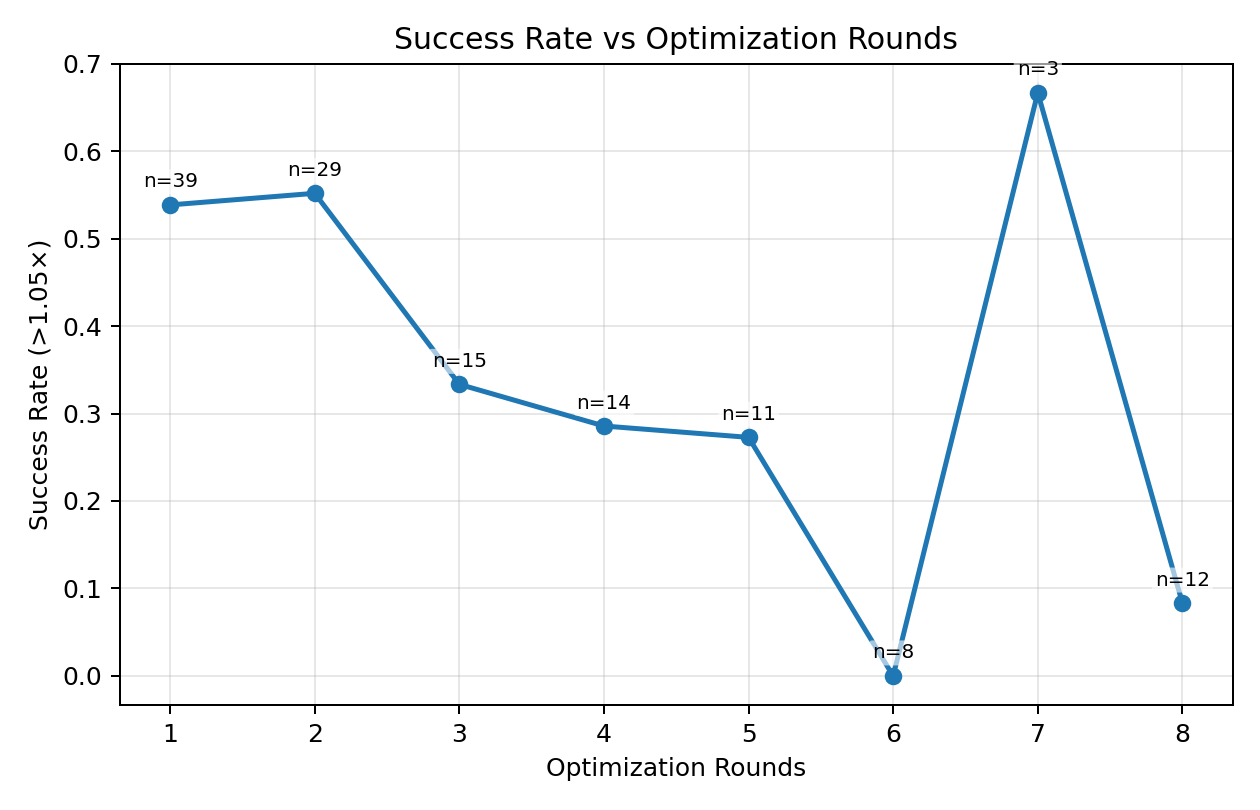}
      \caption{Success rate (and success numbers) vs rounds.}
    \end{subfigure}
  
    \vspace{1em}
  
    \begin{subfigure}[t]{0.495\textwidth}
      \includegraphics[width=\linewidth]{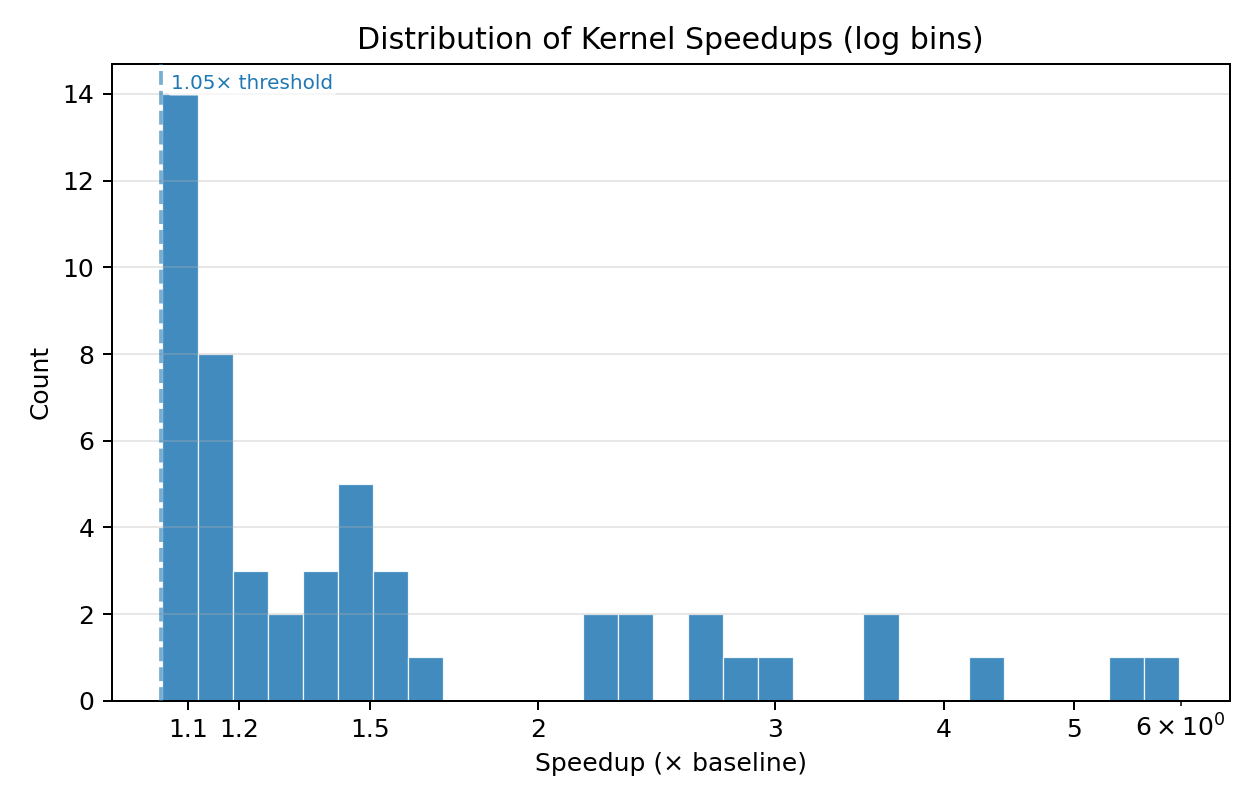}
      \caption{Distribution of speedups.}
    \end{subfigure}
    \hfill
    \begin{subfigure}[t]{0.495\textwidth}
      \includegraphics[width=\linewidth]{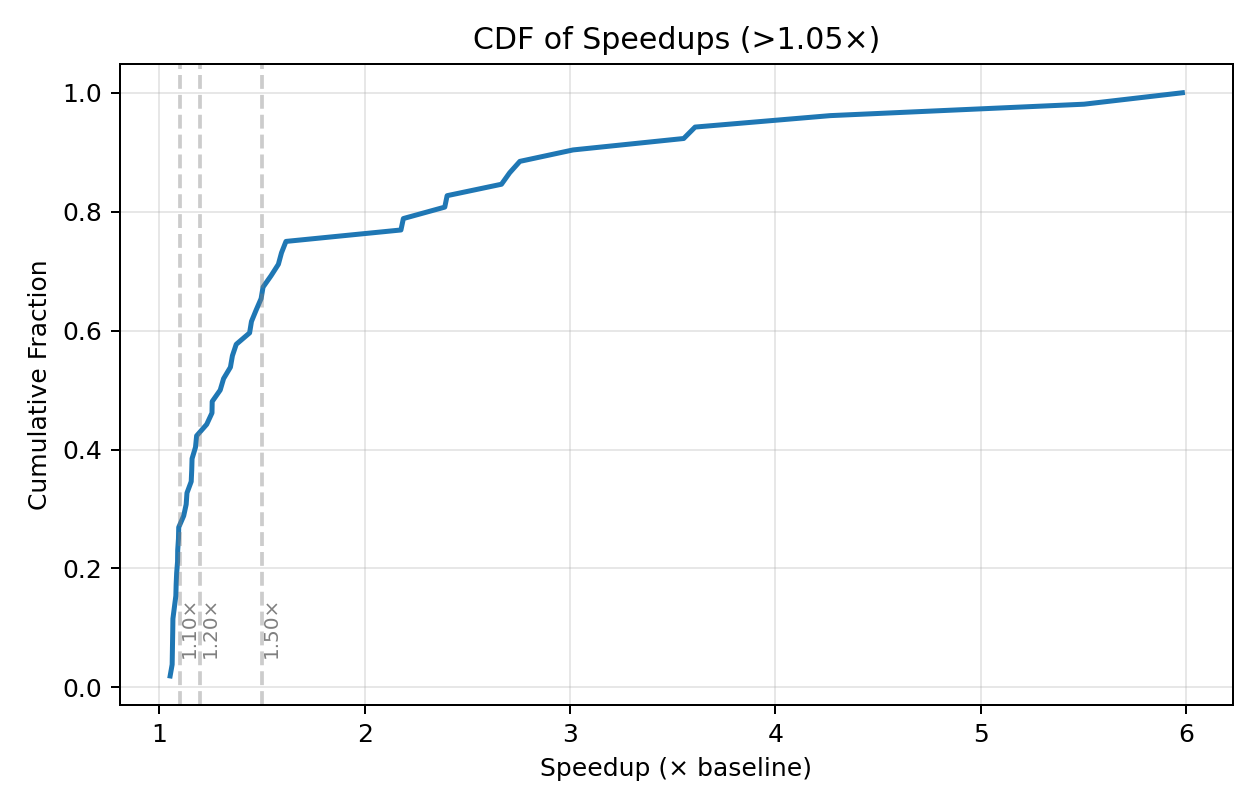}
      \caption{Cumulative distribution (CDF).}
    \end{subfigure}
  
    \caption{
      Overall results of \work.
      (a) The relationship between code length and speedup.
      (b) The relationship between success rate (and number of instances) and number of optimization rounds.
      (c) The distribution of speedups. Most successful kernels achieve 1-1.75x speedups.
      (d) About 40\% (out of successful ones) of kernels reach $\geq$ 1.2x speedup.
    }
    \label{fig:performance_analysis}
\end{figure*}

\subsection{Experimental Setup}

\textbf{Hardware Configuration.} Our experiments are conducted on NVIDIA H100
GPUs with 96GB memory for high-end data center workloads. The system is equipped
with sufficient CPU and memory resources to support the LLM inference and kernel
compilation processes.

\vspace{0.5em}
\noindent
\textbf{Software.} We use Python 3.12 with PyTorch 2.8+ and Triton
3.4 for kernel development and execution. The LLM inference is supported by
Gemini-2.5-Pro as the primary model. NVIDIA Nsight Compute 2025.2.1.0 is used for
performance profiling and analysis. 

\vspace{0.5em}
\noindent
\textbf{Cost Analysis.} We evaluate both the temporal and monetary costs
associated with our multi-iteration workflow. Due to the iterative
nature of \work, resource consumption varies by kernels. On average,
each kernel requires 3.2 optimization rounds, resulting in a mean wall-clock
time of 22 minutes per kernel. The average API cost incurred per kernel is
\$1.20.



\subsection{Datasets and Benchmark}

\textbf{TritonBench.} We evaluate our framework on the complete TritonBench
dataset~\cite{li-etal-2025-tritonbench}, consisting of 184 real-world Triton
kernels collected from GitHub repositories. The dataset covers diverse
computational patterns including matrix operations, convolution kernels,
attention mechanisms, and custom computational kernels, providing a
comprehensive testbed for our optimization approach. Among the 184 kernels,
\work runs on 131 of them (the rest 53 kernels are skipped due to compilation
errors or other issues.) 

\vspace{0.5em}
\noindent
\textbf{Subset for Ablation Studies.}
From the complete dataset of 184 kernels (from tritonbench), we focus to analyze a
subset in detail. To ensure representativeness, we stratify kernels primarily by
their lines of code (LOC), which serves as a proxy for kernel complexity.
Kernels are divided into quantile bins (e.g., Q1–Q4) for the central 90\%
(P5–P95) of the LOC distribution, while very short and very long kernels are
treated as “tail groups.”

From the total, we select 36 kernels: 30 from the central bins (proportionally
allocated across bins) and 6 from the tails (3 shortest and 3 longest). Within
each stratum, kernels are chosen at random with a fixed random seed to guarantee
reproducibility.

This design balances efficiency with coverage: it ensures that kernels of
different sizes and structural complexity are represented, while still keeping
the analysis workload manageable. Results are reported both for the central
population (typical kernels) and for the overall set (central + tails, with
weighting), so that conclusions reflect both mainstream and edge-case behavior.

\vspace{0.5em}
\noindent
\textbf{Overall Results.}
We define a kernel as successful if it achieves a speedup of at least 1.05x
(5\%) compared to the baseline implementation. 
Otherwise, it is considered as failed.
Table~\ref{tab:detailed_results} presents the overall results of \work on the 131
runable kernels. The overall success rate is 42.7\%, the average speedup is 1.76x (on successful kernels).
Table~\ref{tab:detailed_results} also shows the results by kernel length category,
it shows that for whatever kernel length category, their success rate are close to each other.
But the Q1 category achieves the highest average speedup, which is 2.25x.
Case study 1 and 2 (\S\ref{sec:case_study_1} and \S\ref{sec:case_study_2})
show examples of Q1, detailing the optimization opportunities and the optimization process.

\begin{figure*}[t!]
    \centering
    \begin{subfigure}[t]{0.33\textwidth}
        \includegraphics[width=\linewidth]{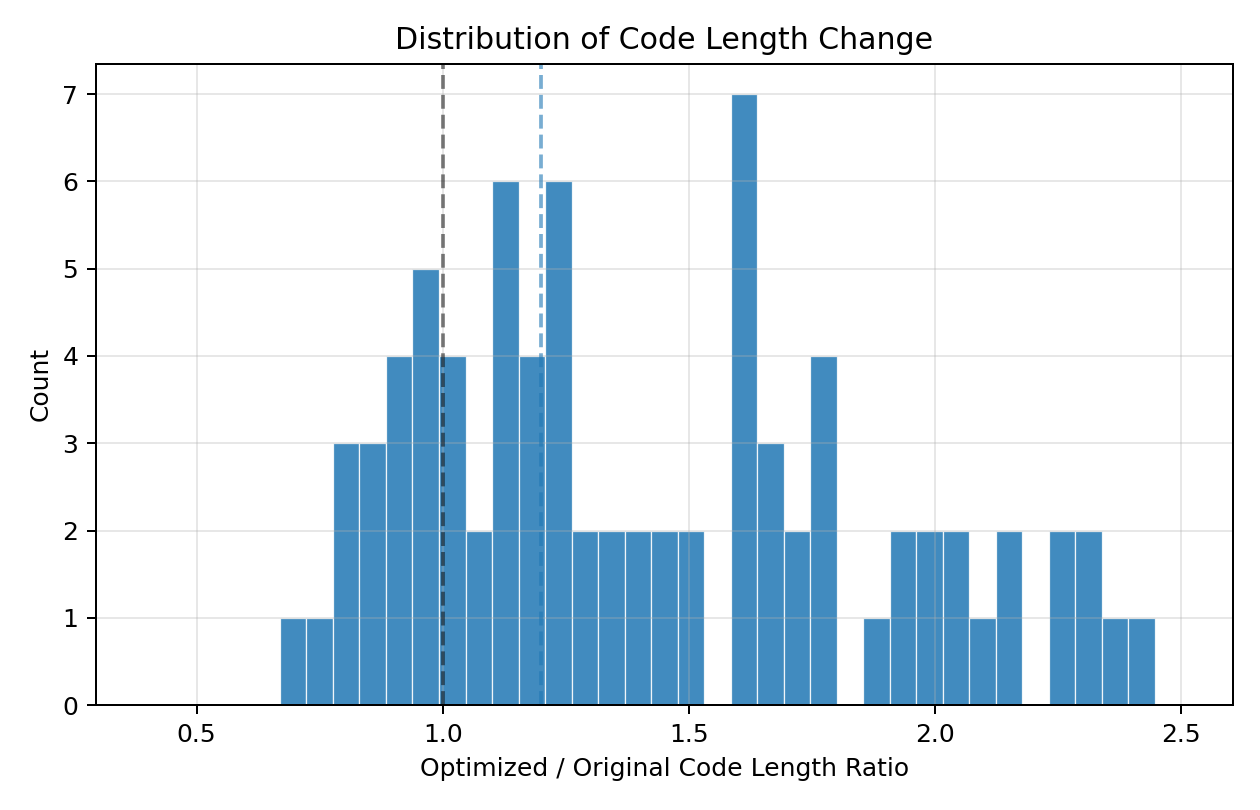}
        \caption{Distribution of code length change.}
    \end{subfigure}
    \hfill
    \begin{subfigure}[t]{0.33\textwidth}
        \includegraphics[width=\linewidth]{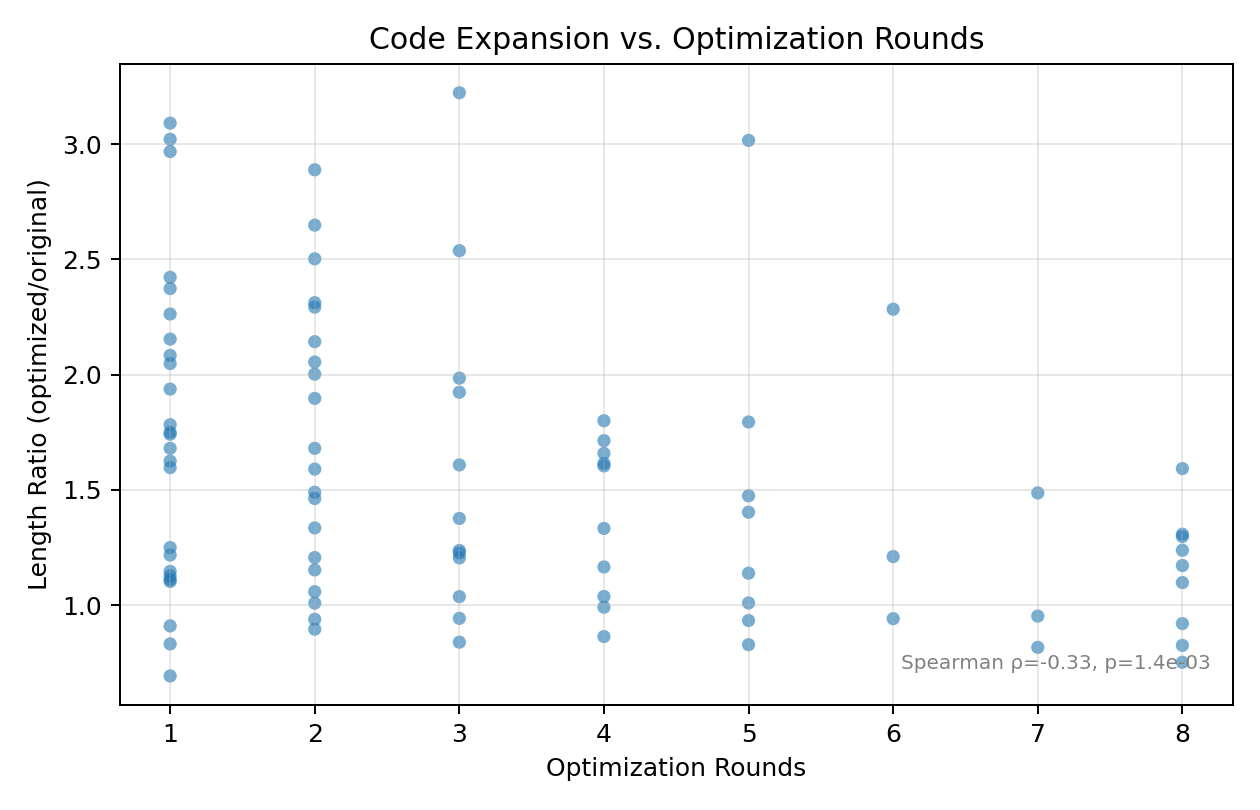}
        \caption{Length ratio vs. optimization rounds.}
    \end{subfigure}
    \hfill
    \begin{subfigure}[t]{0.33\textwidth}
        \includegraphics[width=\linewidth]{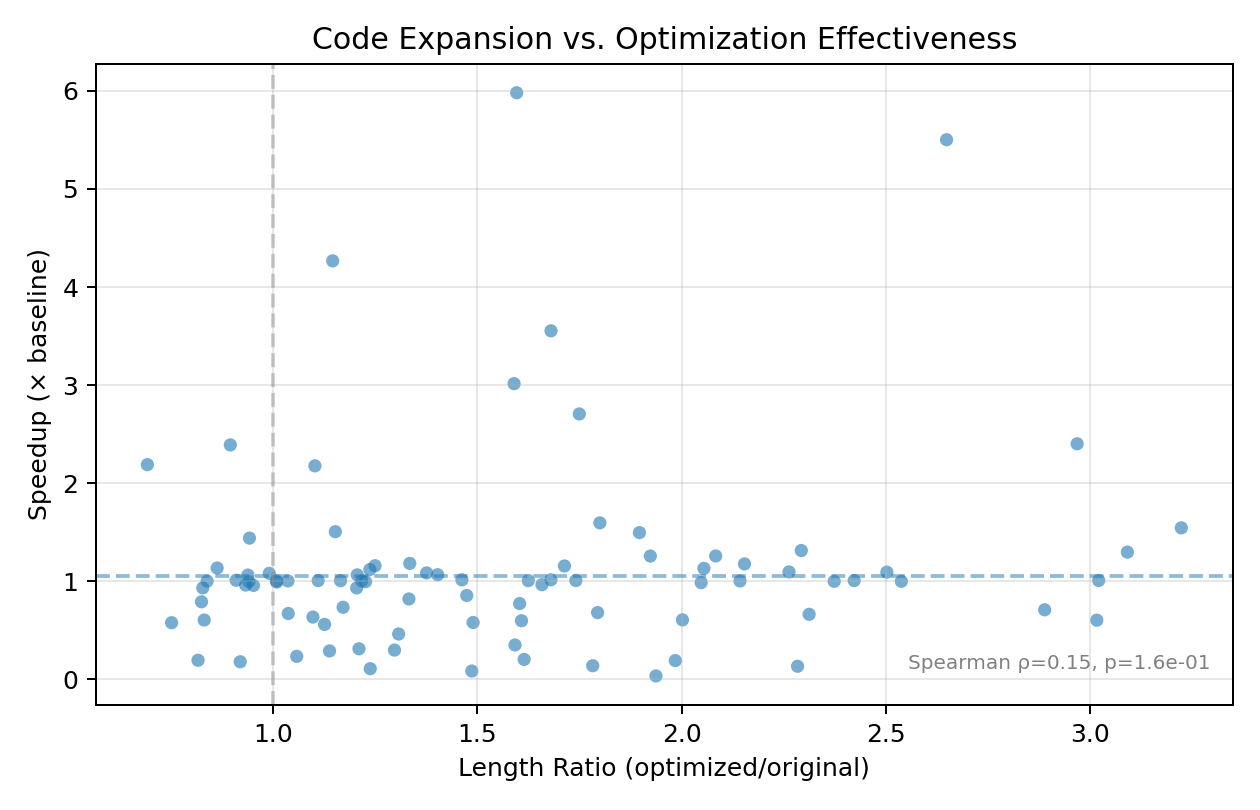}
        \caption{Length ratio vs. speedup.}
    \end{subfigure}

    \caption{
        Analysis of LLM-generated kernel code length.
        (a) The majority of optimized kernels are longer than their originals, 
        indicating a tendency toward code expansion.
        (b) Longer optimization cycles are associated with slightly more compact code.
        (c) Expansion magnitude shows little correlation with performance gain,
        implying that verbosity is not a reliable indicator of actual optimization.
    }
    \label{fig:length_behavior}
\end{figure*}



\begin{table}[htbp]
    \centering
    \caption{Performance Results by Kernel Length Category}
    \label{tab:detailed_results}
    \begin{tabular}{lccc}
    \toprule
    \textbf{Cat.} & \textbf{\# Kernels} & \textbf{Success Rate} & \textbf{Avg. Speedup} \\
    \midrule
    Q1 & 26 & 42.3\% & 2.25x  \\
    Q2 & 35 & 40.0\% & 1.40x \\ 
    Q3 & 35 & 42.8\% & 1.62x \\ 
    Q4 & 22 & 45.5\% & 2.05x \\  
    Tail High & 4 & 50\% & 1.16x \\
    Tail Low & 9 & 66.7\% & 1.79x \\
    \midrule
    Overall & 131 & 42.7\% & 1.76x  \\
    \bottomrule
    \end{tabular}%
\end{table}




\subsection{Performance Characterization}
Figure~\ref{fig:performance_analysis} summarizes the overall performance
behavior of the LLM-based kernel optimization process.
Figure~\ref{fig:performance_analysis}(a) shows the relationship between the
original code length and achieved speedup. While improvements are observed
across various kernel sizes, there is no clear correlation between code
complexity and performance gain, indicating that model-driven optimization is
not constrained by kernel length alone.

In Figure~\ref{fig:performance_analysis}(b), the success rate is plotted against
the number of optimization rounds. A clear downward trend emerges: earlier
rounds achieve higher success rates, whereas excessive iterations often lead to
degradation. This suggests that once the model fails to find meaningful new
strategies, additional refinement attempts may reinforce suboptimal behaviors.

Figure~\ref{fig:performance_analysis}(c) depicts the distribution of observed
speedups (with outliers beyond 10x removed). Most kernels achieve modest
acceleration between 1.05x and 1.5x, while large improvements ($>2x$) are
relatively rare. Finally, the cumulative distribution in
Figure~\ref{fig:performance_analysis}(d) further highlights that about 60\% of
successful cases exceed 1.2x speedup, but fewer than 40\% surpass 1.5x. Overall,
these results demonstrate that the LLM can discover measurable performance
improvements for a subset of kernels, but success is neither uniform across
complexity levels nor guaranteed with repeated optimization attempts.


\subsection{Code Behavior of LLM-Generated Kernels.}
Figure~\ref{fig:length_behavior} summarizes the behavior of code length during 
\work.
As shown in Figure~\ref{fig:length_behavior}(a), most optimized kernels are noticeably longer than their original versions,
with a median length ratio of approximately 1.3$\times$.
This observation suggests that the model generally tends to \textit{expand} or restructure kernels rather than compress them,
possibly introducing redundant control structures or additional intermediate computations
to ensure correctness and readability.

Interestingly, as illustrated in Figure~\ref{fig:length_behavior}(b),
we observe a mild negative correlation ($\rho = -0.33$) between the number of optimization rounds
and the resulting length ratio. This indicates that iterative refinement tends to
reduce the verbosity of earlier generations, leading to more compact yet functionally equivalent code.

Finally, Figure~\ref{fig:length_behavior}(c) compares the degree of code expansion
with the achieved runtime speedup. The correlation is weak and statistically insignificant
($\rho = 0.15$, $p = 0.16$), implying that larger code size does not necessarily translate to better performance.
Together, these findings suggest that the LLM’s inclination toward verbosity is primarily stylistic,
rather than functionally or performance driven.


\begin{table}[t!]
    \centering
    \caption{Profiling result of LLM-generated kernel.}
    \label{tab:profiling_result}
    \begin{tabular}{lcc}
    \toprule
    Performance factor & Before & After \\
    \midrule
    Memory throughput & 52.24\% & \textbf{90.76\%} \\
    Compute throughput & 5.92\% & \textbf{15.00\%} \\
    L2 Cache Throughput & 50.04 \% & \textbf{71.52 \%} \\
    Kernel runtime & 849.50 $\mu$s & \textbf{488.96 $\mu$s} \\
    \bottomrule
    \end{tabular}
    \vspace{-8pt}
\end{table}

\subsection{Ablation Studies}

We conduct comprehensive ablation studies to analyze the individual contributions
of different components in our optimization framework. 
Beyond the complete \work, we also evaluate the following two settings:
(1) LLM E2E (w/o Kernel Execution), and 
(2) LLM One-shot (w/o Iterative Optimization).


\vspace{0.5em}
\noindent
\textbf{LLM E2E (w/o Kernel Execution).} This ablation removes the 
runtime profiling component from \work, evaluating the impact
of performance metrics on optimization quality. In this setting,
the LLM agent relies solely on static code analysis features without access
to detailed profiling information such as memory bandwidth utilization,
compute efficiency, and hardware-specific bottlenecks. This comparison
quantifies the value of runtime profiling data in guiding optimization
decisions and helps assess whether static analysis alone is sufficient
for effective kernel optimization.

\vspace{0.5em}
\noindent
\textbf{LLM One-shot (w/o Iterative Optimization).} This ablation
eliminates the iterative optimization loop, providing the LLM with only the
original kernel code and generating optimized versions in a single pass without
any intermediate kernel execution or performance feedback. This setting
evaluates the LLM's ability to perform optimization based purely on code
understanding and static analysis, without the benefit of iterative refinement
or runtime performance guidance. This comparison helps quantify the contribution
of the iterative optimization process and performance-driven refinement to
overall optimization effectiveness.




\begin{table}[t!]
\centering
\caption{Ablation Study Results on TritonBench Subset (36 kernels)}
\label{tab:ablation_results}
\resizebox{0.48\textwidth}{!}{%
\begin{tabular}{lccccl}
\toprule
\multirow{2}{*}{Method} & \multicolumn{2}{c}{\textbf{Avg. Speedup}} & \textbf{Success Rate} & \textbf{Iterations} \\
 & \textbf{Overall} & \textbf{on Success} & & \\
\midrule
\work & \text{0.96×} & \text{1.51×} & \textbf{42.3\%} & 3.8 \\
LLM E2E & \text{0.95×} & \text{1.33×} & 22.8\% & 3.0 \\
\midrule
LLM One-shot Prompt & \text{1.19×} & \text{2.91×} & 11.4\% & 1.0 \\
\bottomrule
\end{tabular}%
}
\vspace{-8pt}
\end{table}

Table~\ref{tab:ablation_results} presents the comprehensive results of our
ablation studies across the 36 selected kernels from TritonBench. The One-shot
Prompt achieves the best speedup, but only success in very few cases, these
cases are usually simple kernels and optimized by \texttt{autotune} (cases like
\S\ref{sec:case_study_1}). LLM E2E achieves higher success rate to 22.8\%,
showing that LLM can locate the optimization opportunities by itself but still
limited. The \work achieves the best sucess rate to 42.3\%, showing 
the importance of the profiling.


\begin{table*}[ht!]
    \centering
    \caption{Representative case studies illustrating different optimization behaviors of the LLM.}
    \label{tab:case_summary}
    \resizebox{0.98\textwidth}{!}{%
    \begin{threeparttable}
    \begin{tabular}{ccllcl}
        \toprule
        \textbf{Case} &
        \textbf{Cat.} &
        \textbf{Mechanism} &
        \textbf{Type} &
        \textbf{Speedup} &
        \textbf{Key Theme} \\
        \midrule
        \textbf{1} &
        Q1 &
        Introduced autotuning space &
        Meta-level optimization &
        \textbf{95$\times$} &
        Unlocking hidden configuration space \\

        \textbf{2} &
        Q1 &
        Rewrote dataflow via profiling feedback &
        Algorithmic restructuring &
        \textbf{1.74$\times$} &
        Profiling-guided reasoning \\

        \textbf{3} &
        Q3 &
        Repeated low-impact code refactoring &
        Failed optimization &
        $\approx$1.00$\times$ &
        Optimization saturation despite iteration \\
        \bottomrule
    \end{tabular}
    \end{threeparttable}
    }
    
\end{table*}

\section{Case Study}

As shown in Table~\ref{tab:case_summary}, we select three case studies to
illustrate the effectiveness of \work. The first two cases
demonstrate the ability of \work to 
(1) leverage autotuning space to find the optimal configuration and 
(2) restructure the algorithm to improve performance.
The last case demonstrates the limitation of \work when the optimization process
is stuck when facing complex constraints even after multiple optimization rounds.












\subsection{Case Study 1: Autotuning-Driven Optimization}
\label{sec:case_study_1}
This case comes from a simple element-wise operation kernel, called \texttt{sin\_kernel}
The baseline kernel was a minimal mplementation that used a fixed
\texttt{BLOCK\_SIZE=4}, leading to extremely poor GPU utilization and memory
throughput.  
Despite being functionally correct, it effectively serialized the element-wise
computation, yielding a very low baseline performance.

The LLM-optimized version introduced an \texttt{@triton.autotune} decorator with
multiple candidate configurations:  
\begin{lstlisting}[language=Python, caption={Core optimization in the generated kernel.}]
@triton.autotune(
 configs=[
  ...
  triton.Config({'BLOCK_SIZE': 512}, num_warps=4),
  triton.Config({'BLOCK_SIZE': 1024}, num_warps=8),
  triton.Config({'BLOCK_SIZE': 4096}, num_warps=16),
 ],
 key=['n_elements'],
)
@triton.jit
def sin_kernel(...):
 ...
\end{lstlisting}

Although the arithmetic logic remained unchanged, the addition of compile-time
tuning parameters allowed Triton’s autotuner to explore optimal launch
configurations automatically.  
The resulting kernel achieved a remarkable \textbf{95$\times$ speedup} compared
to the naive baseline.  
We \textit{excluded} this case from global statistics (so we claim our best speedup is 5x instead of 95x)
 since its baseline was abnormally weak,
while it still demonstrates that the LLM can identify missing
optimization hooks and leverage the autotuner to find the optimal configuration.

\definecolor{ForestGreen}{RGB}{34,139,34}   
\definecolor{BrickRed}{RGB}{178,34,34}      
\definecolor{GrayText}{gray}{0.35}          
\definecolor{BlueLink}{RGB}{25,70,150}      

\newcommand{\upg}{\textcolor{ForestGreen}{$\uparrow$}}     
\newcommand{\dng}{\textcolor{BrickRed}{$\downarrow$}}      
\newcommand{\neut}{\textcolor{GrayText}{$\rightarrow$}}    
\newcommand{\good}[1]{\textcolor{ForestGreen}{#1}}         
\newcommand{\bad}[1]{\textcolor{BrickRed}{#1}}            

\begin{table*}[!ht]
\centering
\resizebox{0.98\textwidth}{!}{%
\begin{threeparttable}
    \caption{Performance evolution of \texttt{\_bmm\_chunk\_bwd\_kernel} over five optimization rounds on NVIDIA H100.
    Total time includes kernel launch and host-side preprocessing. Arrows indicate change relative to the previous round.}
    \label{tab:kernel-evolution}
    \setlength{\tabcolsep}{5pt}
    \begin{tabular}{clccccccc}
    \toprule
    \textbf{Round} & \textbf{Main Change} &
    \textbf{Total Time} & \textbf{Speedup} &
    \textbf{Kernel Dura.} &
    \textbf{Memory Thru.} & \textbf{SM Thru.} & \textbf{Occupancy} \\
    \midrule
    1 & Baseline (original kernel)
      & 9.69 & 1.00$\times$ & 4.18 & 71.6\% & 52.8\% & 12.5\% \\
    
    2 & Host-side transpose of \texttt{dout}
      & \bad{10.09 \dng} & \bad{0.96$\times$ \dng} & 4.84 & \bad{56.5\% \dng} & \bad{34.3\% \dng} & 12.5\% \\
    
    3 & Smaller tiles \& fewer pipeline stages
      & \bad{10.62 \dng} & \bad{0.91$\times$ \dng} & 6.24 & \bad{43.9\% \dng} & \bad{26.6\% \dng} & 18.8\% \upg \\
    
    4 & Full coalesced rewrite (deep pipeline)
      & \bad{11.23 \dng} & \bad{0.86$\times$ \dng} & 10.6 & 44.0\% \neut & 27.0\% \neut & 18.8\% \\
    
    5 & Occupancy-aware autotune (no transpose)
      & \good{\textbf{9.66} \upg} & \good{\textbf{1.00$\times$} \neut} & \good{3.98 \upg} &
        68.9\% \upg & 41.8\% \upg & \good{\textbf{25.0\%} \upg} \\
    \bottomrule
    \end{tabular}
\end{threeparttable}
}
\end{table*}

\subsection{Case Study 2: Profiling-Guided GEMV Restructuring}
\label{sec:case_study_2}
\textbf{Problem.}
Given a batched vector–matrix multiplication task requiring $2MNK$ FLOPs (where
$M$, $N$, and $K$ denote the number of rows, columns, and batch dimension,
respectively), the goal is to compute:
\[
C[m,n] = \sum_{k=0}^{K-1} A[m,k] \cdot B[m,n,k].
\]

\noindent
\textbf{Baseline Kernel \& Bottleneck.}
The original Triton kernel
performed a 3D broadcast prior to reduction:
\begin{lstlisting}[language=Python]
expanded_a, _ = tl.broadcast(a, b)
acc += tl.trans(tl.sum(expanded_a * b, axis=2))
\end{lstlisting}
This expanded $a$ into a temporary
$[\texttt{block\_n}, \texttt{block\_m}, \texttt{block\_k}]$ tensor,
introducing redundant data movement and a costly transpose stage.

\vspace{0.5em}
\noindent
\textbf{Profiling results} from \texttt{Nsight Compute} revealed:
\squishlist
\item \textbf{Memory throughput:} 52.24\% of theoretical peak
\item \textbf{Compute throughput:} 5.92\% peak $\Rightarrow$ ALUs under-utilized
\item \textbf{Occupancy:} 37.5\% (6 warps per scheduler out of 16)
\item \textbf{Tail effect:} partial wave launches caused up to 50\% runtime waste
\squishend
Overall, execution was clearly memory-bound and dominated by intermediate expansion.

\vspace{0.5em}
\noindent
\textbf{\work Optimization.}
The LLM restructured the kernel into a GEMV-style streamed reduction,
replacing broadcast-based expansion with tile-wise accumulation:
\begin{lstlisting}[language=Python]
acc[n0:n0+Nt] += tl.sum(b_tile * a_tile[None, :], axis=1)
# fp16/bf16 automatically mapped to tl.dot => Tensor Cores
\end{lstlisting}
This formulation enables contiguous access along the $k$-dimension
and exploits hardware dot–product units automatically.

\vspace{0.5em}
\noindent
\textbf{Profiling comparison.}
Table~\ref{tab:profiling_result} summarizes the improvements achieved by the
\work-generated kernel:
\squishlist
\item Removed 3D broadcast $\Rightarrow$ no expanded intermediates
\item Contiguous access along $k$ $\Rightarrow$ improved memory coalescing
\item Lower register pressure $\Rightarrow$ higher occupancy
\item Tunable grid size $\Rightarrow$ mitigated partial-wave tail effect
\squishend

\noindent
\textbf{Outcome.}
The optimized kernel achieved a \textbf{1.74$\times$ speedup},
raising DRAM throughput from 52\% to over 90\% of peak and improving SM
utilization by 2.5$\times$. Profiling confirmed that the model correctly
identified and resolved the dominant memory bottleneck.

\subsection{Case Study 3: Iterative Kernel Optimization}
\label{sec:case_study_3}

Table~\ref{tab:kernel-evolution} summarizes the five iterative optimization rounds applied to the
backward batched-matrix-multiplication kernel.
This kernel computes the gradient term
\[
    \mathbf{dB} = \mathbf{dO}^\top \mathbf{A},
\]
where $\mathbf{A}$ and $\mathbf{dO}$ are chunked along the sequence dimension.
The optimization journey highlights how mutliple 
apparently reasonable optimization choices may impact total time performance
in different ways (and sometimes in opposite directions).

\vspace{0.5em}
\noindent
\textbf{Round 1 (Baseline).}
The baseline kernel performed an in-kernel transpose of \texttt{dout}, yielding non-coalesced memory access but
moderate utilization (kernel = 4.18~ms, total time = 9.69~ms).  
Nsight Compute identified 12.5\% occupancy limited by register pressure.

\begin{lstlisting}[language=Python]
dout_ptrs = dout_ptr + (
    offs_m[:, None] * stride_dout_csize_n +
    offs_cs[None, :] * stride_dout_csize_m
)
\end{lstlisting}

\noindent
\textbf{Round 2 (Host-side transpose).}
A host-side pre-transpose was introduced to enforce coalesced access:
\begin{lstlisting}[language=Python]
dout = dout.transpose(-1, -2).contiguous()
\end{lstlisting}
Although the memory pattern became fully coalesced,
the additional copy and altered layout disrupted cache locality.
The total time latency increased by $\approx$4\%.

\vspace{0.5em}
\noindent
\textbf{Round 3 (Tile-size tuning).}
We reduced the tile dimensions and pipeline stages to lower register usage:
\begin{lstlisting}[language=Python]
triton.Config({'BLOCK_SIZE_M': 64, 'BLOCK_SIZE_N': 128,
               'BLOCK_SIZE_CS': 32}, num_stages=2, num_warps=4)
\end{lstlisting}
Theoretical occupancy rose to 18.8\%, yet smaller tiles under-utilized tensor cores,
increasing launch overhead and slowing runtime to 10.6~ms.

\vspace{0.5em}
\noindent
\textbf{Round 4 (Full coalesced rewrite).}
A full rewrite introduced deep pipelining (\texttt{num\_stages=4}) and complete coalescing.
Despite perfect memory alignment, the kernel consumed $\sim$140~KB of shared memory per CTA,
limiting execution to one block per SM (18.8\% occupancy) and yielding the slowest run (11.2~ms).

\vspace{0.5em}
\noindent
\textbf{Round 5 (Occupancy-aware autotuning).}
We reverted to the original logic but constrained the autotuner to low-shared-memory configurations:
\begin{lstlisting}[language=Python]
triton.Config({'BLOCK_SIZE_M': 64, 'BLOCK_SIZE_N': 256,
               'BLOCK_SIZE_CS': 32}, num_stages=3, num_warps=8)
\end{lstlisting}
These kept shared memory below 76~KB, enabling 3–4 CTAs/SM.
Occupancy increased to 25\% and kernel duration decreases;
but total time speedup is only 1.0028x, thus classified as \emph{no improvement}.

\vspace{0.5em}
\noindent
\textbf{Discussion.}
Across five rounds, the kernel exhibited an instructive pattern: local
improvements in kernel-level metrics did not always translate to total time
gains. The persistent bottleneck was low occupancy from shared-memory and
register pressure. Ultimately, maintaining multiple resident CTAs per SM proved
more impactful than enforcing perfect coalescing.


\section{Limitations \& Discussions}

\textbf{Iterative Optimization Dynamics.}
From both the ablation study and the correlation between optimization rounds
and success rate (Figure~\ref{fig:performance_analysis}),
one might conclude that the LLM should stop early rather than engage
in prolonged refinement loops. 
However, this interpretation can be misleading.
Many of the `early-stopping' successes correspond to simple kernels
where a one-shot prompt is already sufficient for improvement—%
these cases form a small, easier subset of the benchmark.

Increasing the number of refinement rounds allows the LLM to tackle
more complex kernels that cannot be solved in a single step.
The diminishing aggregate success rate therefore reflects the growing difficulty
of newly attempted cases, not merely the inefficiency of later rounds.
Nevertheless, our current iterative design still exhibits low exploration efficiency:
the model often revisits semantically similar variants without making progress.
Future work should focus on improving the search process itself,
for example by incorporating adaptive stopping criteria,
diversity-driven re-prompting, or memory-augmented iterative reasoning.

\vspace{0.5em}
\noindent
\textbf{Partial Effectiveness and Convergence Limits.}
Although \work outperforms the baseline in terms of optimization coverage 
(achieving measurable speedups in roughly 34\% of kernels compared to 11\% for a simple static baseline),
the overall effectiveness remains limited. 
As observed in Case~\#3, the LLM frequently becomes trapped in semantically-equivalent but 
performance-neutral code variants.
This suggests that the model’s exploration space is shallow—once obvious structural fixes are exhausted,
it struggles to identify deeper algorithmic or scheduling opportunities.
The absence of strong gradient-like feedback from profiling signals further limits convergence.

\section{Conclusion}

In this work, we present \work, a novel approach that harnesses the power of
LLMs to automate Triton kernel optimization. \work addresses the challenges
of manual optimization by providing runtime profiling feedback into the code
generation process and shows that LLM-assisted optimization can achieve
performance improvements compared to the baseline implementations.


\bibliographystyle{ACM-Reference-Format}
\bibliography{example_paper}



\end{document}